\documentclass[journal]{IEEEtran}
\usepackage{cite}

\usepackage{graphicx}

\usepackage[cmex10]{amsmath}
\usepackage{amsthm}
\usepackage{amssymb}
\usepackage{cases}
\usepackage{bm}

\usepackage[caption=false,font=footnotesize]{subfig}

\usepackage{fixltx2e}

\usepackage{url}

\usepackage{algorithm}
\usepackage{algorithmic}
\usepackage{xcolor}

\begin{document}
	\title{Channel Fingerprint Based Beam Tracking for Millimeter Wave Communications}
	

\author{Ruichen~Deng,
		Sheng~Chen,
		Sheng~Zhou,~\IEEEmembership{Member,~IEEE,}
		Zhisheng~Niu,~\IEEEmembership{Fellow,~IEEE,}
		Wei~Zhang,~\IEEEmembership{Fellow,~IEEE}
			\thanks{This work was supported by the Nature Science Foundation of China (No. 91638204, No. 61871254, No. 61861136003), the Australian Research Council's Projects funding scheme under LP160100672 and LP160101244, the Tsinghua University Initiative Scientific Research Program, and Hitachi Ltd. \emph{(Corresponding author: Sheng Zhou)}}
			\thanks{R. Deng, S. Chen, S. Zhou and Z. Niu are with Beijing National Research Center for Information Science and Technology, Department of Electronic Engineering, Tsinghua University, Beijing 100084, China. Emails: \{drc13,chen-s16\}@mails.tsinghua.edu.cn,  \{sheng.zhou,niuzhs\}@tsinghua.edu.cn.}
			\thanks{W. Zhang is with the School of Electrical Engineering and Telecommunications, the University of New South Wales, Sydney, NSW 2052, Australia. Email: w.zhang@unsw.edu.au.}
		}

	\maketitle
	\theoremstyle{plain}
	\newtheorem{theorem}{Theorem}
	\theoremstyle{remark}
	\newtheorem{re:re1}{Remark}
	
	\begin{abstract}
	Beamforming structures with fixed beam codebooks provide economical solutions for millimeter wave (mmWave) communications due to the low hardware cost. However, the training overhead to search for the optimal beamforming configuration is proportional to the codebook size. To improve the efficiency of beam tracking, we propose a beam tracking scheme based on the channel fingerprint database, which comprises mappings between statistical beamforming gains and user locations. The scheme tracks user movement by utilizing the trained beam configurations and estimating the gains of beam configurations that are not trained. Simulations show that the proposed scheme achieves significant beamforming performance gains over existing beam tracking schemes.	
	\end{abstract}
	\IEEEpeerreviewmaketitle

	\section{Introduction}
	By providing a large amount of unoccupied spectrum, millimeter wave (mmWave) is an effective solution to satisfy the growing demand of mobile data traffic. Narrow mmWave beams can be formed by antenna arrays to enable high data rate transmission as well as reducing inter-user interference. Considering the hardware cost and the energy efficiency, hybrid and analog beamforming systems using phased arrays are more economical compared with fully digital structures \cite{Health16}. Accurate beam alignment is essential to realize the beamforming gains of phase arrays. With the increasing size of beamforming codebooks, the beam training process will consume a large ratio of channel resources, which can be even larger when users are with high mobility.
	
	Many research efforts have been devoted to designing high-efficient beam training schemes, such as using configurable beam width for adaptive beam search \cite{Alkhateeb14}, sending pseudo-random beacons to apply compressive sensing techniques\cite{Marzi16}, {double-link beam tracking to overcome the blockage problem\cite{Gao14}, probabilistic beam tracking for hybrid beamforming architectures\cite{Palacios17}, adaptive beam tracking with the unscented kalman filter\cite{larew19}} and narrowing down the search range with the historical training results \cite{patra2015smart}. Channel fingerprints can also act as useful historical information to aid beam tracking. The channel fingerprint describes the long-term multi-path channel gains associated with user locations. On one hand, its forward mapping, from channel gains to user locations, is widely used in localization methods \cite{Vo16}. On the other hand, its reverse mapping, from user locations to channel gains, is considered to guide the beam alignment \cite{Va18}, which can greatly improve the beamforming performance, especially on the corner of obstacles. {However, the location information used in these references requires external localization equipments such as GPS devices, which are costly and power-consuming }.
	
	In this work, we propose a channel fingerprint based beam tracking scheme without any additional localization equipments. The localization is realized by the beam training process. Based on the beam training results, the scheme utilizes the forward mapping of the channel fingerprints to estimate the user locations. After that, the reverse mapping of the channel fingerprints is used to estimate the gains of beam configurations that have not been trained. Based on the channel fingerprints, we can improve the quality of beam configurations selected for training and hence increase the beamforming gains compared to existing beam tracking schemes under the same training budget.


	\section{System Model}
	
	The letter focuses on a mmWave beamforming system, where a base station (BS) equipped with a phased array of $M_t$ antennas conducts beamforming to a user equipped with an array of $M_r$ antennas. The user moves in a 2-dimensional rectangle area $\mathcal{A}$. The whole area is discretized into $X=L_1\times L_2$ disjoint rectangle sub-areas, where $L_1$ and $L_2$ are the length and the width of the area, respectively. Each sub-area is assumed to be small enough so that the channel within it is approximately a constant. The time domain is also discretized into frames, whose interval is $\Delta t$. Between frames, the user follows a linear movement model:
	\begin{equation}
	\bm{x}_t = \bm{x}_{t-1}+\bm{v}\Delta t+\bm{n}_W,
	\end{equation}
	where $\bm{x}_t, \bm{x}_{t-1}, \bm{v}\Delta t, \bm{n}_W$ are integer vectors. $\bm{x}_t=[x_t^{(1)},x_t^{(2)}]$ is the coordinate of user location at the $t$-th frame, $\bm{v}$ is the velocity vector measured by the user, and $\bm{n}_W$ is the measurement error, which follows a zero-mean discretized Gaussian distribution:
	\begin{equation}
	p_W(\bm{n}_W)=\int_{\mathcal{N}(\bm{n}_W)} \frac{1}{2\pi |\bm{\Sigma}_W|^\frac{1}{2}}e^{-\frac{\bm{n}\bm{\Sigma}^{-1}\bm{n}}{2}}d\bm{n},
	\end{equation}
	where $\mathcal{N}(\bm{n}_W)=\left\{\bm{n}\vert n^{(i)}-n_W^{(i)}|\leq \frac{1}{2},~~i=1,2\right\}$ is the neighborhood of $\bm{n}_W$.
	
	The system adopts a fixed beamforming codebook $\mathcal{C}$, which consists of beam configurations from the transmitter codebook $\mathcal{C}_t$ and the receiver codebook $\mathcal{C}_r$, i.e., $\mathcal{C}=\{[\bm{w}_{i_t},\bm{v}_{i_r}]|\bm{w}_{i_t}\in\mathcal{C}_t,\bm{v}_{i_r}\in\mathcal{C}_r\}$. Therefore, the size of the codebook is $M=|\mathcal{C}_t||\mathcal{C}_r|$. Using the $i$-th beam configuration $[\bm{w}_{i_t},\bm{v}_{i_r}]$ in $\mathcal{C}$, the beamforming gain is
	\begin{equation}
	\gamma_i = |\bm{v}_{i_r}^H \bm{H} \bm{w}_{i_t}|^2,
	\end{equation}	
	{where $\bm{H}$ denotes the channel matrix with its entry $\mathrm{H}_{i j}$ representing the channel coefficient from the $j$-th transmit antenna to the $i$-th receive antenna. }
	The beamforming gain is characterized by a two-state model. More specifically, the beamforming gain is (in dB scale)
    \begin{equation}
	    \gamma_i^\text{NDB}(\bm{x},t)=g_i^\text{NDB}(\bm{x})+n_{V}^\text{NDB}(t)
    \end{equation}
    in the non-dynamic-blockage state, and
    \begin{equation}
	    \gamma_i^\text{DB}(\bm{x},t)=g_i^\text{DB}(\bm{x})+n_{V}^\text{DB}(t)
    \end{equation}
    in the dynamic-blockage state, where $i$ is the beamforming configuration index, $t$ is the frame index, and $\bm{x}$ is the user location. $g_i^\text{NDB,DB}(\bm{x})$ and $n_{V}^\text{NDB,DB}(t)$ stand for the beamforming gain component due to the large and small scale channel fading, respectively.

	Compared to microwave, mmWaves have weaker ability in the aspect of scattering and reflecting and are more susceptible to the dynamical blockage events caused by obstacles such as vehicles and human bodies. Therefore, the beamforming gain component $g_i^\text{DB}(\bm{x})$ tends to 0. Meanwhile, we assume $n_{V}^\text{NDB}(t)$ and $n_{V}^\text{DB}(t)$ follow the same Gaussian distribution with zero mean and variance $\sigma_V^2$. The occurrence of the non-dynamic-blockage state is modeled as a Bernoulli variable $\delta_i(\bm{x},t)$, i.e., $\delta_i(\bm{x},t)=1$ represents the non-dynamic-blockage state, whose probability is $\alpha$, and $\delta_i(\bm{x},t)=0$ represents the dynamic-blockage state. Therefore, the two equations can be integrated into one:
	\begin{equation}
	\gamma_i(\bm{x},t)=\delta_i(\bm{x},t)g_i(\bm{x})+n_V(t).
	\end{equation}
	The dynamic blockages are assumed to be independent and identically distributed (i.i.d.) over frame index $t$ and beam configuration $i$ \footnote{ While occurrences of blockage in successive frames are correlated, this temporal correlation is actually beneficial to the proposed scheme for reducing the uncertainty of  beamforming gains. Our work simplifies the temporal correlation part of the beam model, and the results can be further improved by integrating the temporal statistics of beamforming gains.}. Compared to the varying small scale fading, the noise in the beam training process has a much smaller variance. Therefore, we ignore the training noise and use the training result $\hat{\gamma}_i$ to approximate the real beamforming gain $\gamma_i$. The training budget, i.e., the number of beam configurations trained in each frame, is denoted by $T$.

	\section{Channel Fingerprint Based Beam Tracking}
	\begin{figure}[!t]
		\centering
		\includegraphics[width=3.5 in]{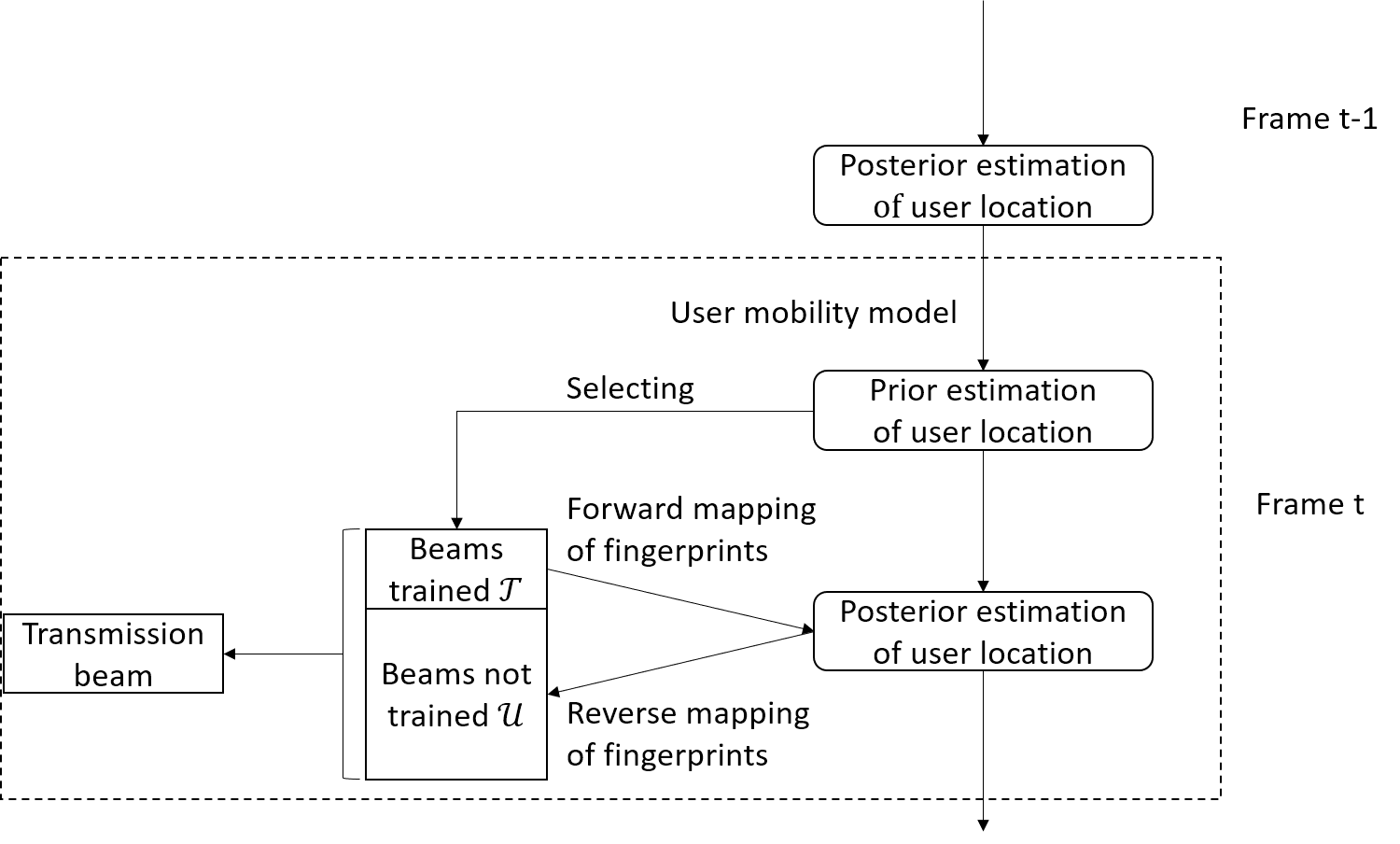}
		\caption{The flow of the channel fingerprint based beam tracking scheme.}
		\label{fig:TrackingScheme}
	\end{figure}
	We describe the proposed channel fingerprint based beam tracking scheme in this section, whose flow is depicted in Fig. \ref{fig:TrackingScheme}. The scheme tracks beam configurations frame-by-frame. At the beginning of each frame, the scheme obtains the prior estimation of the user location by combining the user mobility model and the posterior estimation in the last frame. Based on the prior estimation of the user location, a subset $\mathcal{T}$ of the codebook is selected for beam training. Then the posterior estimation of the user location is determined by the training results and the prior estimation. After that, the beam configurations not trained is estimated through the reverse mapping of the fingerprint database $\bm{g}(\bm{x})$. Finally, the scheme selects the beam configuration with the highest estimated gain for data transmission. Different from many existing beam tracking schemes such as \cite{Va16BT}, our proposed scheme keeps updating user positions instead of directly tracking beam directions. The reason is twofold: tracking user positions can better utilize the channel fingerprints, and meanwhile it can be easily integrated with other localization methods to improve the performance. In each frame, the user location is estimated by combining the beam training results and the user mobility model. After that, the beamforming gains get estimated based on the estimated user location. More specifically, we consider two concrete beam tracking algorithms, namely the one based on the recursive Bayesian estimation (RBE) and the other one based on the extended Kalman filter (EKF). The former focuses on the distribution estimation of user locations, while the latter focuses on the point estimation of user locations.
	\subsection{Beam Tracking Based on the Recursive Bayesian Estimation}
	The initial probability distribution of user location is denoted by $p_{0|0}(\bm{x})$, where the symbol $t_1|t_2$ represents the estimation or the distribution of the location at the $t_1$-th frame conditioned on the beam training results at the $t_2$-th frame and the frames before. If the distribution $p_{0|0}(\bm{x})$ is not given in advance, the scheme may use a priori assumptions, such as the uniform distribution in the whole area $\mathcal{A}$.
	
	At the beginning of each frame, the prior probability distribution of user location is derived based on the probability distribution in the last frame and the transition probability in the mobility model:
	\begin{equation}\label{apriori_loction}
	p_{t|t-1}(\bm{x}) = \sum_{\bm{n}_W} p_{t-1|t-1}(\bm{x}-\bm{v}\Delta t-\bm{n}_W)p_W(\bm{n}_W).
	\end{equation}
	
	The beam configurations for training is selected based on the prior location distribution $p_{t|t-1}(\bm{x})$. The optimal choice should consider both the benefits in the current frame and the frames forth, which can be formulated as a Markov decision process (MDP). The state, action and reward of the MDP are the prior location distribution, the beam configurations selected for training and the beamforming gain in the data transmission process, respectively. Obtaining the solution to the MDP needs huge computation efforts since there are infinite states. Therefore, we consider the myopic solution, which is suboptimal but computationally efficient\cite{MDPbook}. More specifically, the scheme selects the $T$ beam configurations with the largest expected gains for training:
	\begin{equation}\label{trainBeamSelection_RBE}
	\mathcal{T}=\arg\max_i(T)\left\{\alpha \sum_{\bm{x}} g_i(\bm{x}) p_{t|t-1}(\bm{x})\right\},
	\end{equation}
	where $\arg\max_i(T)\{\cdot\}$ stands for the $T$ indexes with the largest values in the set.
	
	Then the prediction $p_{t|t-1}$ is refined by the beam training results according to the Bayes' rule:
	\begin{equation}\label{posterior_RBE}
	p_{t|t}(\bm{x}) = \frac{f(\hat{\bm{\gamma}}(t)|\bm{x})p_{t|t-1}(\bm{x})}{\sum_{\bm{x}'}f(\hat{\bm{\gamma}}(t)|\bm{x}')p_{t|t-1}(\bm{x}')},
	\end{equation}
	where the likelihood is calculated as
	\begin{equation}
	f(\hat{\bm{\gamma}}\!(t)\!|\bm{x}) \!=\! \frac{1}{(2\pi)^{\!\frac{T}{2}\!}\sigma_V^T}\!\prod_{i=1}^{M}\!\left(\!\alpha e^{\!-\!\frac{\vert\hat{\gamma}_i(t)\!-\!g_i(\bm{x})\vert^2}{2\sigma_V^2}}\!+\!(1\!-\!\alpha)e^{\!-\!\frac{\vert\hat{\gamma}_i(t)\vert^2}{2\sigma_V^2}}\!\right)
	\end{equation}
	The refined result becomes the posterior probability distribution of the user location in the $t$-th frame. The estimated user location can thus be determined as the expectation of the posterior probability distribution:
	\begin{equation}
	\hat{\bm{x}}_{t|t}=\sum_{\bm{x}}\bm{x} p_{t|t}(\bm{x}).
	\end{equation}
	
	For the beam configuration not trained $i\in\mathcal{U}$, we use the prior probability distribution for the blockage estimation. Therefore, the expectation of the beamforming gain is
	\begin{equation}\label{EBG_U}
	\hat{\gamma}_i(t)=\alpha \sum_{\bm{x}} g_i(\bm{x}) p_{t|t}(\bm{x}),~~i\in\mathcal{U}.
	\end{equation}
	
	For the beam configuration $i\in\mathcal{T}$, we use the training result $\hat{\gamma}_i(t)$ as the estimation. After obtaining the estimated gains of the whole beam configurations, the configuration with the largest gain is selected for the data transmission process:
	$i^\star_t = \arg\max_{i\in\mathcal{C}} \hat{\gamma}_i(t)$.
	The procedure of the RBE based beam tracking scheme is summarized in Algorithm \ref{alg:RBE}.
	\begin{algorithm}[t]
		\setlength{\baselineskip}{12pt}
		\caption{The RBE based beam tracking scheme}\label{alg:RBE}
		\begin{algorithmic}[1]
			
		\REQUIRE ~~                          
		Initial location distribution $p_{0|0}(\bm{x})$, user velocity $\bm{v}$, the distribution of the location measurement error $p_W(\bm{n}_W)$ and the fingerprint database $\bm{g}(\bm{x})$.
		\ENSURE ~~                           
		The index of transmit beam configuration $i_t^\star,~~t=1,2,\cdots$.
		\FOR{$t=1,2,\cdots$}
			\STATE Obtain the prior location distribution $p_{t|t-1}(\bm{x})$ according to (\ref{apriori_loction}).
			\STATE Select the beam configurations for training according to (\ref{trainBeamSelection_RBE}) and get the training results $\hat{\bm{\gamma}}(t)$.
			\STATE Obtain the posterior location distribution $p_{t|t}(\bm{x})$ according to (\ref{posterior_RBE}).
			\STATE Estimate the gains of the beam configurations not trained according to (\ref{EBG_U}), and select the beamforming configuration $i_t^\star$ with the largest gain for data transmission.
		\ENDFOR
		\end{algorithmic}
		
	\end{algorithm}
	\subsection{Beam Tracking Based on the Extended Kalman Filter}
	The EKF algorithm extends the Kalman filter to non-linear cases, which suits our tracking problem. The EKF based scheme generates point estimation of user locations instead of the prior and posterior distributions generated in the RBE based scheme. Therefore, it is less accurate than the RBE based scheme, but saves a lot of computational complexity.
	
	At the beginning of each frame, the prior estimation $\hat{\bm{x}}_{t|t-1}$ is obtained as:
	\begin{equation}\label{prior_EKF}
	\hat{\bm{x}}_{t|t-1}=\hat{\bm{x}}_{t-1|t-1}+\bm{v}\Delta t.
	\end{equation}
	
	Similar to the RBE based scheme, the EKF based scheme selects the $T$ beam configurations with the largest expected gains at location $\hat{\bm{x}}_{t|t-1}$ for training:
	\begin{equation}\label{trainBeamSelection_EKF}
	\mathcal{T}=\arg\max_i(T)\left\{\alpha  g_i(\hat{\bm{x}}_{t|t-1})\right\}.
	\end{equation}
	
	The beam training results are $\hat{\bm{\gamma}}(t)$. Different from the RBE framework, where the estimate of the dynamic blockage is implicitly included in the calculation of the posterior distribution of user locations, we need to explicitly estimate the dynamic blockage of beam configurations trained in the EKF based scheme, which is done by the maximum a posteriori (MAP) estimation algorithm. More specifically, the dynamic blockage is determined by comparing with a threshold:
	\begin{equation}\label{blockageEstimate}
	\hat{\delta}_{i}(t)=\begin{cases}
	1 & \hat{\gamma}_i(t)>\frac{1}{2}g_i(\hat{\bm{x}}_{t|t-1})-\frac{\sigma_V^2\ln\frac{\alpha}{1-\alpha}}{g_i(\hat{\bm{x}}_{t|t-1})},\\
	0 & \hat{\gamma}_i(t)\leq \frac{1}{2}g_i(\hat{\bm{x}}_{t|t-1})-\frac{\sigma_V^2\ln\frac{\alpha}{1-\alpha}}{g_i(\hat{\bm{x}}_{t|t-1})}.
	\end{cases}
	\end{equation}
	
	If the estimation $\hat{\delta}_{i}(t) = 0$ for all the beam configurations trained (which comes with very small probability when the training budget is enough), then the training results provide no useful information for the user location and the system uses the prior estimation $\hat{\bm{x}}_{t|t-1}$ as the posterior estimation. Otherwise, the system chooses the training results with $\hat{\delta}_{i}(t) = 1$ to form the effective vector $\gamma^E(t)$. Then the channel fingerprint gradient at the prior location for each beam configuration in $\gamma^E(t)$ is calculated as
	\begin{equation}
	\nabla g_i(\hat{\bm{x}}_{t|t-1})) = \left(\frac{\partial g_i(\hat{\bm{x}}_{t|t-1})}{\partial \hat{\bm{x}}_{t|t-1}^{(1)} },\frac{\partial g_i(\hat{\bm{x}}_{t|t-1})}{\partial \hat{\bm{x}}_{t|t-1}^{(2)} }\right)^T,
	\end{equation}
	where the superscripts $(i), i=1,2$ represents the two spatial dimensions. All the gradients form a gradient matrix $\bm{J}$. The Kalman gain is then obtained:
	\begin{equation}
	\bm{K}_g = \bm{P}_{t|t-1}\bm{J}^T(\bm{J}\bm{P}_{t|t-1}\bm{J}^T+\sigma_V^2\bm{I})^{-1},
	\end{equation}
	where the prior error covariance is
	\begin{equation}
	\bm{P}_{t|t-1} = \bm{P}_{t-1|t-1}+\sigma_W^2\bm{I}.
	\end{equation}
	
	After that, the posterior estimation of the user location is derived:
	\begin{equation}\label{posterior_EKF}
	\hat{\bm{x}}_{t|t}=\hat{\bm{x}}_{t|t-1}+\bm{K}_g(\gamma^E(t)-g^E(t)),
	\end{equation}
	
	Based on the posterior location estimation, the system gets the estimation
	\begin{equation}\label{posteriorBeamformingGain_EKF}
	\hat{\gamma}_i(t)=\alpha g_i(\hat{\bm{x}}_{t|t}),~~i\in\mathcal{U},
	\end{equation}
	for those beams without training. Finally, the beam with the highest estimated gain is selected for transmission.
	
	The procedure of the EKF based beam tracking scheme is summarized in Algorithm \ref{alg:EKF}.
	
	\begin{algorithm}[t]
		\setlength{\baselineskip}{12pt}
		\caption{The EKF based beam tracking scheme}\label{alg:EKF}
		\begin{algorithmic}[1]
			
			\REQUIRE ~~                          
			Initial location estimation $\hat{\bm{x}}_0$, user velocity $\bm{v}$, and the fingerprint database $\bm{g}(\bm{x})$.
			\ENSURE ~~                           
			The index of transmit beam configuration $i_t^\star,~~t=1,2,\cdots$.
			\FOR{$t=1,2,\cdots$}
			\STATE Obtain the prior location estimation $\hat{\bm{x}}_{t|t-1}$ according to (\ref{prior_EKF}).
			\STATE Select the beam configurations for training according to (\ref{trainBeamSelection_EKF}) and get the training results $\hat{\bm{\gamma}}(t)$.
			\STATE Estimate the dynamic blockage $\hat{\delta}_{i}(t)$ according to (\ref{blockageEstimate}).
			\IF{$\hat{\delta}_{i}(t)=0$ for all $i\in\mathcal{T}$}
			\STATE	Obtain the posterior location estimation as $\hat{\bm{x}}_{t|t}=\hat{\bm{x}}_{t|t-1}$.
			\ELSE
			\STATE Obtain the posterior location estimation according to (\ref{posterior_EKF}).
			\ENDIF
			\STATE Estimate the gains of the beam configurations not trained according to (\ref{posteriorBeamformingGain_EKF}), and select the beamforming configuration $i_t^\star$ with the largest gain for data transmission.
			\ENDFOR
		\end{algorithmic}
		
	\end{algorithm}
	\section{Simulation Results}
	The simulation scenario is a 2-dimensional urban street with length 100 m and width 4 m, where the channel data is generated by the ray-tracing results of the \emph{Wireless Insite} software\cite{WirelessInsite}. The area is divided into sub-areas of $0.1$ m by $0.1$ m. The time interval of a frame is $\Delta t = 20 $ ms.The velocity estimation is $\bm{v}=[v,0]$ (which means the user moves along the street horizontally with velocity $v$). The covariance matrix of the measurement error vector is
	$$\bm{\Sigma} = \begin{bmatrix}
	\frac{1}{2}\sigma_W^2 & 0\\
	0&\frac{1}{2}\sigma_W^2
	\end{bmatrix},$$ which represents the case where the measurement errors of the two dimensions are independent and with the variation $\frac{1}{2}\sigma_W^2$. The default value of $v$ and $\sigma_W$ are set to $15$ m/s and $1$, respectively. On the other hand, the BS uses a uniform linear array with 64 antenna elements to conduct beamforming for the user. The array operates at frequency 28 GHz and the antenna elements are separated by half-wavelength. We use progressive phase shifter to generate the beamforming vector. Hence the transmit codebook size is 64. The user is equipped with a single antenna, making the receive codebook size 1. Therefore the size of the total codebook is 64. The simulation results are averaged over 1000 runs. In every simulation run, the initial location of the user is set to the coordinate $[1,20]$, and then beam tracking algorithms are performed in the next 100 frames. {For simplicity, we assume that the initial location is accurate for both the RBE and EKF schemes, based on the observation that the impact of initial location estimation vanishes with the growing number of frames.} {We consider the \emph{beamforming gain gap} $\Delta \gamma$ and the training coverage ratio $p_\text{TC}$ as the performance metrics. The former is defined as the average of the gap between the beamforming gain of tracking algorithms and the groundtruth of the highest beamforming gain:
	\begin{equation}
	\Delta \gamma(t) = \max_i\gamma_i(t)-\gamma_{i_t^\star}(t),
	\end{equation}
	and the latter is defined as the ratio of the frames in which the highest beamforming configuration is trained:
	\begin{equation}
	p_\text{CT} = \frac{\text{card}\{t|\arg\max_i\gamma_i(t)\in\mathcal{T}(t)\}}{\text{card}\{t\}},
	\end{equation}
	where $\text{card}\{\cdot\}$ denotes the cardinality of the set.}

\begin{figure}[!t]
		\centering
		\includegraphics[width=3 in]{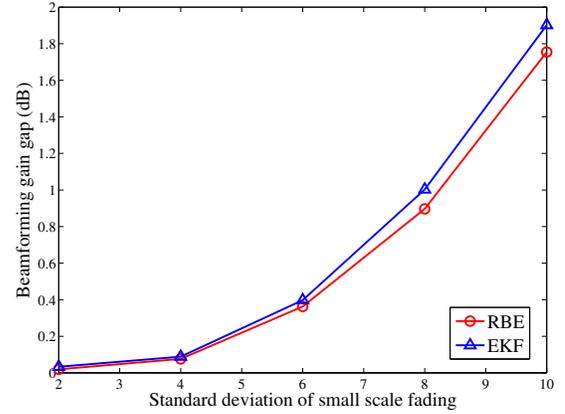}
		\caption{The relationship between the beamforming gain gap and the standard deviation of small scale fading. The number of beam configurations trained is $T=5$ and the probability of dynamic blockage is $1-\alpha=0.2$.}
		\label{fig:BG_sigmaV}
	\end{figure}

\begin{figure}[!t]
	    \centering
	    \includegraphics[width=3 in]{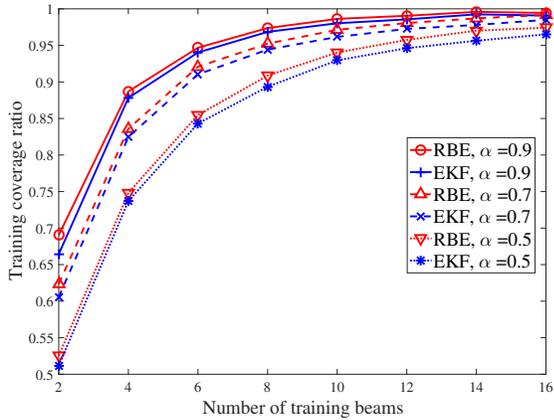}
	    \caption{{The relationship between the training coverage ratio and the number of training beams. The standard deviation of small scale fading is $\sigma_V=6$ dB. }}
	    \label{fig:PC_N_alpha}
	\end{figure}

	The performance of the RBE and EKF tracking algorithms is shown in Fig. \ref{fig:BG_sigmaV} and Fig. \ref{fig:PC_N_alpha}. {The standard deviation $\sigma_V$ of small scale fading increases with the growing extent of multi-path propagation.} We can observe that the performance of RBE and EKF decreases with $\sigma_V$. Besides, the proposed algorithm has an obvious performance advantage over EKF and the advantage grows with $\sigma_V$. The reason is that the proposed algorithm preserves a better characterization of user location for the next frame than EKF.
	
	{On the other hand, the training coverage ratios of both algorithms increase with the growing number of training beams as well as the decreasing blockage probability. When $N$ grows to $10$, even the EKF algorithm with $\alpha=0.5$ (the dynamical blockage probability is half) can achieve a $93\%$ coverage ratio.}
	
\begin{figure}[!t]
			\centering
			\includegraphics[width=3 in]{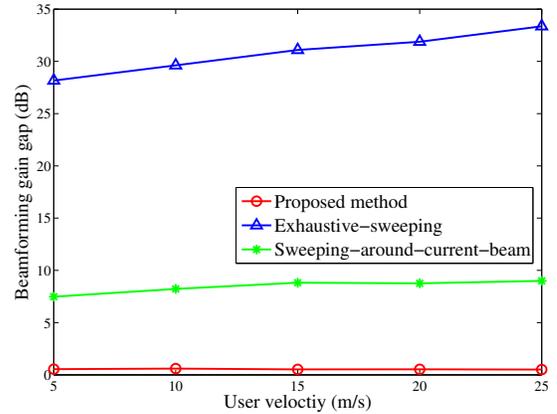}
			\caption{The comparison of different beam tracking schemes. The standard deviation of small scale fading is $\sigma_V=6$ dB and the probability of dynamic blockage is $1-\alpha=0.2$.}
			\label{fig:BG_v}
	\end{figure}

	Finally, we compare the performance of the proposed scheme with other beam tracking schemes in Fig. \ref{fig:BG_v}. {For simplicity, we use the RBE algorithm for the proposed beam tracking scheme.} The number of beam configurations trained in all the schemes is set to $4$. The exhaustive-sweeping scheme conducts training for all the beam configurations. The RBE based beam tracking scheme and the sweeping-around-current-beam scheme are applied in each frame, while the exhaustive-sweeping scheme is applied every $16$ frames to keep the same training overhead with other schemes. In the exhaustive-sweeping scheme, the frame without training selects the same beam configuration as the previous frame. The sweeping-around-current-beam scheme \cite{patra2015smart} searches the beam configurations whose directions are close to the one selected in the last frame (including the selected), which improves the performance of the exhaustive-sweeping by utilizing the historical beam training.
	The correlation of beamforming gains between adjacent frames decreases with the growing user velocity, which worsens the performance of the exhaustive-sweeping scheme and the sweeping-around-current-beam scheme. It is observed that the proposed scheme has the best robustness to user mobility because the fingerprint database provides prior information about the beamforming gains.
	When the user velocity is $25$ m/s, the proposed scheme further reduces the beamforming gain gap by about $8.5$ dB compared to the sweeping-around-current-beam scheme with the aid of the channel fingerprints.
	\section{Conclusion}\label{sec:conclusion}
	In this letter, we have proposed a high-efficiency beam tracking scheme based on the channel fingerprints for mmWave beamforming systems. Under the same training budget, the proposed scheme can improve the selected transmitting beamforming gain by as large as $8.5$ dB for high mobility users compared to the existing sweeping-around-current-beam scheme. Future work will consider optimizing the choice of beam configurations trained to further improve the performance. {Also, the dynamic updating of the channel database with respect to the changes of the environment is also a promising future direction.}
	

	
	\bibliographystyle{IEEEtran}
	\bibliography{myref}

\begin{thebibliography}{10}
\providecommand{\url}[1]{#1}
\csname url@samestyle\endcsname
\providecommand{\newblock}{\relax}
\providecommand{\bibinfo}[2]{#2}
\providecommand{\BIBentrySTDinterwordspacing}{\spaceskip=0pt\relax}
\providecommand{\BIBentryALTinterwordstretchfactor}{4}
\providecommand{\BIBentryALTinterwordspacing}{\spaceskip=\fontdimen2\font plus
\BIBentryALTinterwordstretchfactor\fontdimen3\font minus
  \fontdimen4\font\relax}
\providecommand{\BIBforeignlanguage}[2]{{%
\expandafter\ifx\csname l@#1\endcsname\relax
\typeout{** WARNING: IEEEtran.bst: No hyphenation pattern has been}%
\typeout{** loaded for the language `#1'. Using the pattern for}%
\typeout{** the default language instead.}%
\else
\language=\csname l@#1\endcsname
\fi
#2}}
\providecommand{\BIBdecl}{\relax}
\BIBdecl

\bibitem{Health16}
R.~W. {Heath}, N.~{González-Prelcic}, S.~{Rangan}, W.~{Roh}, and A.~M.
  {Sayeed}, ``An overview of signal processing techniques for millimeter wave
  {MIMO} systems,'' \emph{IEEE Journal of Selected Topics in Signal
  Processing}, vol.~10, no.~3, pp. 436--453, Apr. 2016.

\bibitem{Alkhateeb14}
A.~{Alkhateeb}, O.~{El Ayach}, G.~{Leus}, and R.~W. {Heath}, ``Channel
  estimation and hybrid precoding for millimeter wave cellular systems,''
  \emph{IEEE Journal of Selected Topics in Signal Processing}, vol.~8, no.~5,
  pp. 831--846, Oct 2014.

\bibitem{Marzi16}
Z.~{Marzi}, D.~{Ramasamy}, and U.~{Madhow}, ``Compressive channel estimation
  and tracking for large arrays in mm-wave picocells,'' \emph{IEEE Journal of
  Selected Topics in Signal Processing}, vol.~10, no.~3, pp. 514--527, April
  2016.

\bibitem{Gao14}
B.~{Gao}, Z.~{Xiao}, C.~{Zhang}, L.~{Su}, D.~{Jin}, and L.~{Zeng},
  ``Double-link beam tracking against human blockage and device mobility for
  60-ghz wlan,'' in \emph{2014 IEEE Wireless Communications and Networking
  Conference (WCNC)}, April 2014, pp. 323--328.

\bibitem{Palacios17}
J.~{Palacios}, D.~{De Donno}, and J.~{Widmer}, ``Tracking mm-wave channel
  dynamics: Fast beam training strategies under mobility,'' in \emph{IEEE
  INFOCOM 2017 - IEEE Conference on Computer Communications}, May 2017, pp.
  1--9.

\bibitem{larew19}
S.~G. {Larew} and D.~J. {Love}, ``Adaptive beam tracking with the unscented
  kalman filter for millimeter wave communication,'' \emph{IEEE Signal
  Processing Letters}, vol.~26, no.~11, pp. 1658--1662, Nov 2019.

\bibitem{patra2015smart}
A.~Patra, L.~Simi{\'c}, and P.~M{\"a}h{\"o}nen, ``Smart mm-wave beam steering
  algorithm for fast link re-establishment under node mobility in 60 {GHz}
  indoor {WLANs},'' in \emph{Proceedings of the 13th ACM International
  Symposium on Mobility Management and Wireless Access}.\hskip 1em plus 0.5em
  minus 0.4em\relax ACM, 2015, pp. 53--62.

\bibitem{Vo16}
Q.~D. {Vo} and P.~{De}, ``A survey of fingerprint-based outdoor localization,''
  \emph{IEEE Communications Surveys Tutorials}, vol.~18, no.~1, pp. 491--506,
  Firstquarter 2016.

\bibitem{Va18}
V.~{Va}, J.~{Choi}, T.~{Shimizu}, G.~{Bansal}, and R.~W. {Heath}, ``Inverse
  multipath fingerprinting for millimeter wave {V2I} beam alignment,''
  \emph{IEEE Transactions on Vehicular Technology}, vol.~67, no.~5, pp.
  4042--4058, May 2018.

\bibitem{Va16BT}
V.~{Va}, H.~{Vikalo}, and R.~W. {Heath}, ``Beam tracking for mobile millimeter
  wave communication systems,'' in \emph{IEEE Global Conference on Signal and
  Information Processing}, Dec 2016, pp. 743--747.

\bibitem{MDPbook}
M.~L. Puterman, \emph{Markov decision processes: discrete stochastic dynamic
  programming}.\hskip 1em plus 0.5em minus 0.4em\relax John Wiley \& Sons,
  2014.

\bibitem{WirelessInsite}
Remcom, ``{Wireless Insite},''
  [Online].~~Available:~\url{https://www.remcom.com/wireless-insite-em-propagation-software.}~Accessed
  on: Mar. 2019.

\end{thebibliography}

\end{document}